# Moiré reconstruction at a twisted $SrTiO_3$ interface

Pengcheng Fei[1,‡], Meiyu Wang[1,2,‡], Suzhen Liu[3,‡], Zhongwei Zhang[4,‡], Yang Wu[1,‡], Bo Hao[1,‡], Hanhao Zhang[5], Dingcheng Ma[1], Shijie Yang[1], Weiye Nie[1], Jiacheng Wang[1], Jibo Xu[1], Geliang Shen[1], Junjie Wang[4], Chi Ding[4], Huifeng Chen[1], Sheng Xu[1], Xinrui Yang[1], Lu Han[1,*], Haocheng Quan[1], Ling-Ling Ma[1], Zaiyao Fei[5], Liqi Zhou[6], Songhua Cai[3,*], Zhengbin Gu[1], Yu Deng[1], Jian Sun[4,*], Di Wu[1,*], and Yuefeng Nie[1,*]

[1]National Laboratory of Solid State Microstructures, College of Engineering and Applied Sciences, Jiangsu Key Laboratory of Artificial Functional Materials, and Jiangsu Physical Science Research Center, Nanjing University, Nanjing 210093, P. R. China.
[2]Center for Shared Scientific Research Facilities, Nanjing University, Nanjing 210093, P. R. China.
[3]Department of Applied Physics, The Hong Kong Polytechnic University, Kowloon, Hong Kong SAR, China.
[4]National Laboratory of Solid State Microstructures, School of Physics and Collaborative Innovation Center of Advanced Microstructures, Nanjing University, Nanjing 210093, P. R. China.
[5]National Laboratory of Solid-State Microstructures, School of Electronic Science and Engineering and Collaborative Innovation Center of Advanced Microstructures, Nanjing University, Nanjing 210093, P. R. China.
[6]Institute of Materiobiology, College of Sciences, Shanghai University, Shanghai, 200444, China

[‡]These authors contributed equally: Pengcheng Fei, Meiyu Wang, Suzhen Liu, Zhongwei Zhang, Yang Wu, Bo Hao
*Email: lhan@nju.edu.cn; songhua.cai@polyu.edu.hk; jiansun@nju.edu.cn; diwu@nju.edu.cn; ynie@nju.edu.cn

**Abstract**
Twisted oxide double layers have emerged as a promising platform for exploring moiré physics, including polar topologies and emergent electronic phases. However, the intrinsic complexity of oxide surfaces, such as contamination, roughness, and the presence of two distinct terminations (*A*O and *B*$O_2$), can complicate the identification of buried interfacial structures. Here, using atomically flat $SrTiO_3$ membranes with a predominantly SrO-$TiO_2$ interfacial configuration, we directly visualize a moiré-periodic structural reconstruction at the buried interface that is distinct from the geometric moiré of rigid stacking. The reconstructed interface hosts head-to-head-like polar distortions, moiré-modulated Ti electronic structure, and enhanced microwave admittance in otherwise paraelectric and insulating $SrTiO_3$. Our work demonstrates that, with a well-defined termination, moiré reconstruction could serve as a route to emergent electrical behavior in twisted perovskite oxides.

## 1. Introduction

Moiré patterns arise as quasiperiodic in-plane projections when two identical or similar periodic lattices are stacked with a relative twist. In atomically thin van der Waals (vdW) materials, this geometric effect has given rise to twistronics—a field in which the twist angle serves as a powerful tuning parameter for engineering electronic band structures[1]. Twisted double layer graphene provides a representative example, with phenomena ranging from Hofstadter's butterfly[2-4] to the valley Hall effect[5,6] to magic angle strongly correlated physics[7,8], inspiring a wealth of subsequent discoveries in twisted van der Waals heterostructures[9-11]. More importantly, stacking- and twist-induced interfacial ferroelectricity has emerged in two-dimensional materials, including double layer boron nitride, marginally twisted hexagonal boron nitride, and twisted semiconducting double layers, demonstrating that moiré reconstruction can generate polar states even in systems whose constituent layers are not intrinsically ferroelectric[12-15]. These successes have motivated the extension of moiré engineering to transition metal oxides (TMOs) perovskites. Unlike vdW materials with dangling-bond-free surfaces[16], oxides feature a three-dimensional $ABO_3$ framework with two distinct surface terminations ($A$O and $BO_2$), offering a richer set of stacking configurations and interlayer bonding characteristics (e.g., ionic, covalent, or van der Waals-type, depending on the terminating planes). This structural complexity, combined with diverse electronic phases found in transition-metal oxides, including high-temperature superconductivity, colossal magnetoresistance, and multiferroicity in oxides[17-22], makes twisted oxide double layers a compelling yet underexplored frontier for emergent moiré phenomena beyond 2D systems.

Freestanding perovskite oxide membranes, synthesized via sacrificial layer techniques, serve as versatile building blocks that can be deterministically stacked and twisted[23]. Recent studies have reported emergent phenomena in twisted oxide double layers, including polar-distortion vortices in both ferroelectric and paraelectric systems[24-26], moiré-periodic charge disproportionation[27], and twist-controlled chirality[28]. Recent work[29] however, has shown that plan-view microscopy alone cannot distinguish an atomically bonded interface from two membranes separated by a finite gap. Consequently, the direct correlation among local stacking configuration, three-dimensional atomic reconstruction, electronic redistribution, and emergent functional responses remains largely unexplored. Meanwhile, in twisted vdW double layers below ~1°, lattice relaxation produces uniform stacking domains separated by narrow domain walls[30,31]. In twisted oxide double layers such as $BaTiO_3$, theoretical studies at larger twist angles have also predicted substantial atomic relaxation[32], manifesting as chiral vortex displacement patterns but without the formation of well-defined stacking domains. Direct experimental characterization of such interfacial reconstruction, particularly its three-dimensional atomic structure, remains lacking in twisted oxide systems, as does its connection to macroscopic electrical responses.

Here, we engineer the interfacial termination of twisted $SrTiO_3$ double layers and resolve the reconstructed interface using cross-sectional imaging. Rather than simply observing moiré patterns, our work establishes a systematic correlation between atomic-scale interfacial reconstruction, moiré-modulated polar-distortion and electronic states, and their emergent macroscopic electrical responses. This provides a general framework for understanding and identifying reconstruction-driven moiré functionalities in twisted oxide membranes.

## 2. Results

### 2.1 Predicted moiré-modulated structural reconstruction at the interface

For twisted $SrTiO_3$ double layers with SrO-$TiO_2$ termination, the stacking-energy landscape is described by the generalized stacking fault energy (GSFE)[31-33] as a function of in-plane sliding in Extended Data Fig. 1. This landscape defines three characteristic stacking orders within the moiré supercell: AA sites (bulk-like stacking, Sr on Sr or Ti on Ti) as the energy minimum; AB sites (Sr on Ti) as the energy maximum; and SP (saddle point) regions as the transition zones between different stacking orders along domain walls.

To investigate the structural reconstruction of the twisted interface, we developed a machine learning potential (MLP) based on density functional theory (DFT) calculations (see Methods) and performed molecular dynamics (MD) simulations. Figures 1a and 1b present schematic illustrations of the reconstructed structure. Atoms adjacent to the twisted interface undergo pronounced reconstruction, characterized by the expansion of low-energy AA regions and the contraction of high-energy AB regions. Several unit cells away from the interface, the layers remain rigidly stacked without significant relaxation, forming a three-dimensional structure comprising four distinct horizontal planes: top surface ($Top_{sur}$), top interface ($Top_{inter}$), bottom interface ($Bottom_{inter}$), and bottom surface ($Bottom_{sur}$), as shown schematically in Fig. 1c. Figure 1d schematically depicts the relaxed interface, consistent with the simulated relaxed structure in Fig. 1e. Notably, the four AA-type stackings within a moiré supercell are not all equivalent[34]: the centers of adjacent AA regions are offset by half a lattice constant along the moiré lattice directions ($[100]_M$ or $[010]_M$). We therefore denote these two distinct types as AA and AA'. Figure 1f, which is the cross-sectional projection of the relaxed structure, clearly shows this half-unit-cell offset between the Sr columns. Three-dimensional schematic illustrations of the atomic structures for the AA, AA', AB, and SP regions are provided in Extended Data Fig. 2. We also simulated the reconstruction of SrO-SrO and $TiO_2$-$TiO_2$ interfaces. As shown in Extended Data Fig. 3, all relaxed structures expand in the low-energy region according to the fault energies in Table S1.

### 2.2 Atomically sharp interface and direct visualization of moiré reconstruction

A series of $SrTiO_3$ (STO) films with controlled thicknesses (15, 20, 25 unit cell thickness (u.c.)) were grown by molecular beam epitaxy (MBE)[35,36] on approximately 3.5 u.c. water-soluble $Sr_4Al_2O_7$ sacrificial layers deposited on STO (001) substrates. It enables the release of the freestanding STO film, which was subsequently assembled into twist-controlled STO double layers. Given the SrO-rich, water-labile chemistry of the $Sr_4Al_2O_7$ sacrificial layer and the dynamic surface stoichiometry evolution during co-deposition growth, the released bottom surface of the STO membrane is designed to retain a Sr-rich surface chemistry. For the 25 u.c. STO films, for example, the growth mode was switched to shutter growth[37,38] in the last eight layers. The evolution of reflection high-energy electron diffraction (RHEED) oscillations was monitored to obtain a $TiO_2$-terminated top surface. Thus, the bottom (SrO) and top ($TiO_2$) terminations of the freestanding STO film are independently controlled. The freestanding films were then released by dissolving the $Sr_4Al_2O_7$ sacrificial layer in deionized water and transferred onto substrates. Twist-controlled double layers were assembled by sequential layer-by-layer transfer under optical alignment, yielding a SrO-$TiO_2$ interface configuration ($Top_{inter}$: SrO; $Bottom_{inter}$: $TiO_2$). Subsequent annealing was performed to induce atomic-scale reconstruction at the interface. More details about film growth, transfer, annealing and characterization are provided in the Materials and Methods section and Extended Data Figs. 4 and 5.

Figure 2 compares three distinct interfacial states of the twisted STO double layers. Figures 2a-c and 2d–f show plan-view and cross-sectional HAADF-STEM images, respectively. Figures 2a and 2d show a sample with a clearly resolved interfacial gap. In Fig. 2a, regions exhibiting alternating intensities between adjacent atomic columns are assigned to the AA stacking, whereas regions with nearly uniform column intensities correspond to the AB stacking. At this stage, the AA and AB regions have comparable sizes, and the SP regions connecting them remain nearly square. Figure 2d further reveals a gap of approximately 2 nm between the two membranes[29]. Figures 2b and 2e show a sample with an interface with no obvious gap that nevertheless remains approximately rigidly stacked. Fig. 2b displays features similar to those observed in Fig. 2a. Figure 2e, however, shows that the top and bottom STO layers already form an atomically sharp interface. At this point, the interface exhibits the characteristic structure of the $ABO_3$ perovskite, and no interfacial reconstruction is observed.

Finally, Figures 2c and 2f show the reconstructed sample after annealing at 750 °C for 2 h in an oxygen atmosphere. In Fig. 2c, the AA regions expand substantially, whereas the AB regions contract correspondingly, and the SP regions elongate into rectangles, consistent with the simulated results (Fig. 1e). Through-focus STEM imaging of a reconstructed plan-view sample shows a depth-dependent evolution of the moiré contrast consistent with the theoretically expected interfacial reconstruction (Extended Data Fig. 6). These changes indicate that the interface has deviated from the rigid stacked configuration and undergone pronounced reconstruction. A schematic illustrating the annealing conditions used to prepare the three distinct interfacial states is shown in Extended Data Fig. 7. In the cross-sectional image shown in Fig. 2f, the electron beam is aligned with the moiré lattice along $[100]_M$. Under this imaging condition, neither the top nor the bottom layer is exactly aligned with its bulk lattice axis. The bright interfacial region exhibits an intensity comparable to that of the bulk layers, consistent with atomically intimate contact[29]. The specimens shown in Fig. 2c and Fig. 2f were obtained from different samples transferred onto different substrates. To enable direct comparison between plan-view and cross-sectional measurements at the same twist angle, we repeated the experiment on SiN substrates and obtained a comprehensive set of structural and spectroscopic data (Extended Data Fig. 8). Images of the same area obtained by separately aligning to the lattice axes of the top and bottom layers are provided in Extended Data Fig. 9. Notably, the dark regions at the interface are assigned to the AB regions, which preserve atomic arrangements similar to those in rigidly stacked interfaces with larger twist angles. By comparison, the bright regions correspond to reconstructed AA regions, where neighbouring AA and AA' registries are offset by approximately half a lattice constant. At the interface, alternating Sr- and Ti-dominated projected columns are observed (Fig. 2f), arising from this registry shift and thereby deviating from the bulk-like perovskite projection. The termination dependence and reproducibility of the interfacial reconstruction were further examined in a series of samples. Similar reconstruction was consistently observed in SrO-$TiO_2$-terminated samples across different twist angles. By contrast, local changes in the interfacial termination disrupt the periodic reconstruction and the associated alternating interfacial contrast. Additional data are provided in Extended Data Figs. 10, 11 and 12. Overall, these control experiments provide direct atomic-scale evidence for termination-sensitive, moiré-periodic interfacial reconstruction in twisted SrO-$TiO_2$-terminated $SrTiO_3$ double layers.

### 2.3 Interfacial polar distortion, Ti electronic-state modulation and switchable electromechanical response

The observed interfacial reconstruction motivates an examination of the associated microscopic and macroscopic responses of the system. To examine the local polar-distortion at the atomic scale, we acquired HAADF images with the electron beam aligned along the lattice directions of the top and bottom layers, respectively (Figs. 3a and 3d). The positions of Sr and Ti atomic columns were identified by two-dimensional Gaussian fitting. Because oxygen columns are not reliably resolved by HAADF-STEM, we quantified the cation polar displacement from the position of each Ti column relative to the geometric centre of its four neighbouring Sr columns, starting approximately 1–2 unit cells away from the interface. Born effective charges were used to compare relative local dipole trends rather than to determine absolute polarization values, following established displacement-based approaches[39]. The results reveal head-to-head polar displacements near the interface. In the two representative regions shown (arrows indicate polar-distortion direction and magnitude), the polar-distortion is stronger in the dark (AB) regions and oriented more perpendicular to the interface. Toward the AA regions, the polar-distortion rotates and its magnitude decreases slightly. In addition, the polar-distortion is mainly confined to a few atomic layers near the interface and decays rapidly away from it, in agreement with previous theory work[40]. It should be noted that, in cross-sectional specimens, the AA and AB region signals are not pure due to unavoidable SP contributions along the thickness direction. Thus, our results are more suitable for comparing relative trends than for determining absolute polar-distortion values. Oxygen-sensitive iDPC-STEM measurements were also performed. The oxygen sublattice, however, exhibits a systematic apparent displacement, consistent with the strong sensitivity of iDPC imaging to specimen tilt and residual imaging distortions. Besides, the reconstructed registry produces projection overlap that prevents unambiguous assignment of individual oxygen columns (Extended Data Fig. 13). Complementary DPC measurements show a clear contrast reversal across the reconstructed interface, in contrast, no comparable reversal is observed in the gapped control sample (Extended Data Fig. 14). Together with the HAADF-derived cation displacement, these observations support a head-to-head-like interfacial polar configuration. Cross-sectional geometric phase analysis (GPA) further reveals a periodic moiré-modulated strain distribution at the reconstructed interface, with enhanced strain in the AB regions spatially correlated with the local polar distortion (Extended Data Fig. 15).

To further assess the three-dimensional polar and strain response, we also analysed plan-view polar-distortion and GPA maps before and after reconstruction (Extended Data Figs. 16 and 17). The reconstructed sample exhibits a distortion of the in-plane strain network, consistent with the structural relaxation. By contrast, interpretation of the plan-view polarization maps is less straightforward: owing to the small specimen thickness, signals from the upper and lower layers remain superimposed even when focusing near the interface, giving rise to polarization vortices in both the reconstructed and unreconstructed samples. We therefore regard cross-sectional imaging as more reliable for resolving the interfacial polar configuration in these thin specimens.

Atomistic calculations likewise yield reconstruction-associated three-dimensional polar-distortion textures (Extended Data Figs. 18 and 19), although the detailed out-of-plane configuration of the idealized SrO–$TiO_2$ interface differs from the experimentally observed head-to-head-like pattern. We therefore use the calculations to establish the emergence and spatial modulation of reconstruction-associated polar distortions, rather than their absolute magnitude or exact local topology. For comparison, we also examined the symmetric interfaces, e.g., SrO-SrO and $TiO_2$-$TiO_2$ terminations. In these cases, head-to-head out-of-plane polar-distortion structures are observed in regions with relatively higher generalized stacking fault energy.

To examine the spatial correlation between the local Ti electronic structure and the moiré structure, we performed a large-area STEM-EELS map across the interface (Fig. 3b). Equal-area integration windows were first selected from the top layer, the AA region, the AB region, and the bottom layer, and the Ti-$L_{2,3}$ spectra were extracted with a unified background subtraction and normalization procedure. As shown in Fig. 3c, compared with the top and bottom regions that are closer to bulk-like STO, the Ti $L_3$ main peak in both the AA and AB regions exhibits a clear red-shift, with a larger shift in the AB region. These results indicate an interface-localized modification of the Ti electronic structure that varies with the local moiré registry. Although a red-shift of the Ti $L_3$ main peak at the interface has also been reported in previous work, no obvious interfacial reconstruction was observed[27]. The cross-sectional EELS provides direct depth-resolved information of the buried interface, whereas plan-view EELS involves projection averaging along the beam direction. The interfacial Ti valence is likely governed by the interplay of multiple factors, including changes in Ti coordination and the associated oxygen octahedral distortion[41], by shear strain[42], and even by the introduction of screw dislocations at the interface[43]. Quantitative $Ti^{3+}$/$Ti^{4+}$ fitting and O-K edge analysis do not reveal a clear moiré-periodic signature (Extended Data Fig. 20). These measurements therefore do not support a simple picture in which the periodic Ti $L_3$ shift reflects only a corresponding modulation of the average Ti valence or oxygen-vacancy concentration. Instead, we attribute the robust spectral signature to a moiré-correlated modulation of the local Ti electronic structure arising from the reconstructed interfacial environment, without assigning it to a single microscopic mechanism. This behavior was consistently reproduced across different samples, supporting its association with the reconstructed moiré interface (Extended Data Fig. 21). By contrast, no comparable interfacial spectral modulation is observed in the gapped control sample (Extended Data Fig. 22). To further substantiate this conclusion, in Fig. 3e, we constructed a Ti $L_3_e_g$ map from the 4 × 4 binned STEM-EELS spectrum-imaging dataset, with the energy scale calibrated to the zero-loss peak. The largest peak shift is found in the region corresponding to the dark AB contrast in Fig. 3b; at the same time, the entire interfacial region shows an overall red-shift relative to the top and bottom regions. Finally, in Fig. 3f, line profiles were extracted from the same area in the ADF image and the Ti $L_3_e_g$ map. To reduce fluctuations of the line profile, spatial averaging was performed using an 8 × 16 pixel block during the scan. The red curve represents the ADF intensity profile, whereas the blue curve shows the Ti $L_3_e_g$ peak position. The two exhibit a clear spatial correspondence, further supporting the correlation between local interfacial structural modulation and reconstruction of the local Ti electronic state.

Moiré-induced interfacial reconstruction also gives rise to pronounced changes in the macroscopic response of the system. Figure 4a presents a schematic three-state switching picture, illustrated using a Landau-type double-well energy landscape. Figures 4d and 4e show the PFM responses of the reconstructed twisted STO double layers after box-in-box poling at ±90 V. To minimize electrostatic cross-talk, the data were acquired 60 min after poling. Figure 4d presents the phase image, in which the inner and outer boxes exhibit clear contrast (State II/III), whereas the unwritten region retains a disordered as-grown background (State I). Figure 4e shows the amplitude image, where the written regions display comparable response amplitudes that are markedly enhanced relative to the unwritten area, with a distinct dark domain-wall-like feature at the boundary between them. As a control, the same measurement protocol was applied to a single-layer region of the same sample. As shown in Figs. 4g and 4h, after the same ±90 V writing procedure and a 60-min delay, no phase or amplitude contrast is observed. Local PFM spectroscopy measurements were performed on different twisted STO samples. The phase hysteresis loops and amplitude butterfly loops consistently exhibit voltage-dependent switching characteristics (Extended Data Fig. 23), supporting a switchable electromechanical response in the reconstructed twisted interfaces. Notably, samples on conductive Si substrates exhibited substantially lower switching voltages (Extended Data Fig. 24), compared to the reconstructed 7.2° double-layer on an insulating Si substrate. This Highlights the strong influence of substrate conductivity on the effective switching voltage. Together with the controlled growth stoichiometry and the absence of a switching response in single-layer STO, these controls argue against Sr-rich composition or surface-related effects as the dominant origin of the observed response[44,45]. Similarly, neither the rigidly stacked nor the gapped samples exhibit a discernible PFM contrast after poling with a sufficiently high applied voltage (Extended Data Fig. 25). We therefore infer that, in the paraelectric STO system, interfacial moiré reconstruction induces interfacial polar distortions. The evolution of the double-well energy is closely associated with the local increase in $c/a$ ratio induced by interfacial lattice compression in the AB/SP region[46,47], where the $c/a$ ratio exceeds that of the bulk, corresponding data are provided in the Extended Data Fig. 26. As a result, the unwritten region displays only a weak net ferroelectric-like response, as the oppositely oriented interfacial polar-distortions largely compensate each other. Taken together, the interfacial moiré reconstruction gives rise to polar distortions, whereas electrical poling promotes a more ordered polar configuration, manifested as enhanced PFM contrast.

Since head-to-head polar configurations may promote local charge accumulation and thereby enhance interfacial conduction[48,49], we first illustrated the buried conductive network at the reconstructed interface in twisted STO double layers (Fig. 4b). To probe whether the interfacial transport response is enhanced, scanning microwave impedance microscopy (sMIM) measurements were performed across the monolayer/double layer boundary. Figure 4i shows the surface topography, and Figs. 4j and k presents the imaginary and real-channel sMIM signal acquired from the same area. A clear enhancement of the local microwave response is observed in the double layer region in Fig. 4j, indicating an increased local admittance. To exclude a trivial thickness effect, we prepared another STO double layer sample with the same thickness but an unreconstructed interface; sMIM measurements were also performed in the corresponding monolayer/double layer boundary region, as shown in Figs. 4l, m and n. No comparable signal enhancement is observed in the imaginary-channel image. Comparison with the simultaneously acquired topography further shows that regions of larger surface height do not systematically exhibit stronger sMIM contrast; some elevated regions instead show weaker signals, arguing against simple topographic crosstalk as the origin of the enhanced microwave response. A further before-and-after control on the same twisted sample shows no comparable enhancement before annealing, whereas a pronounced double-layer sMIM response emerges after the annealing treatment that produces the reconstructed interface (Extended Data Fig. 27). Together, these measurements associate the enhanced microwave-admittance response with the reconstructed double-layer interface rather than with layer thickness or surface morphology alone. It is worth noting that, in the 4.62° twisted system, the moiré period is well below the lateral resolution (several tens of nanometers) of conventional PFM[50] and sMIM[51]. Consequently, the measured signals represent a spatially averaged response over the entire double layer region, rather than resolving the local network of AB/SP domains. On this basis, we anticipate that direct observation of the corresponding moiré patterns in PFM and sMIM should become feasible in future small-angle samples with precisely controlled twist angles. In conclusion, coexistence of reconstruction-associated polar distortions, a retained switchable electromechanical response and enhanced interfacial admittance, highlights the potential of this system for multilevel non-volatile memory, memristive neuromorphic devices, and reconfigurable nanoelectronics.

## 3. Discussion

In this work, we have directly visualized moiré-periodic atomic reconstruction at the buried interface of twisted $SrTiO_3$ double layers with a well-defined $SrO$-$TiO_2$ termination using cross-sectional HAADF-STEM. This reconstruction is accompanied by moiré-modulated polar distortions and modulation of the local Ti electronic structure. At larger length scales, the reconstructed double layers exhibit a switchable electromechanical response and an enhanced microwave-admittance response. These emergent properties are absent in non-reconstructed or monolayer controls, confirming that moiré-modulated reconstruction is the key driver of the observed functionalities.

Further reducing the twist angle to increase the moiré period would enable direct imaging of local electromechanical and microwave-admittance responses using PFM and sMIM with improved correspondence to individual moiré stacking regions. The stacking-dependent coupling between polar distortion and Ti electronic-state modulation shows how local moiré registry can organize multiple interfacial degrees of freedom. The associated switchable electromechanical and microwave-admittance responses further suggest opportunities for reconfigurable oxide interfaces. More broadly, our work demonstrates that termination-engineered moiré reconstruction offers a predictable and designable route to emergent electronic phenomena in complex oxides, and provides a framework for distinguishing geometric moiré contrast from reconstruction-driven structural and functional responses.

## Methods

***Theoretical calculations:*** The neuroevolution potential (NEP)[52] was used to build the MLP. The DFT relaxations used for the training data were performed with the Vienna Ab initio Simulation Package (VASP)[53], where the Perdew-Burke-Ernzerhof functional modified for solids (PBEsol) within the generalized gradient approximation was chosen as the exchange-correlation functional[54]. The projector augmented-wave method was used with the kinetic cutoff energy of 800 eV together with a k-mesh spacing of 0.3 $Å^{-1}$[55]. The DFT-D3 method was employed to take the van der Waals (vdW) interaction into account[56]. The energy and force convergence criteria were set to $1 \times 10^{-6}$ eV and $1 \times 10^{-3}$ eV/Å. An active learning framework NEPMaker was used to generate a comprehensive training set of structures with a small twist angle[57]. The trained MLP achieved mean absolute errors of 4 meV per atom for energy and 99 meV $Å^{-1}$ for force on the training set. Error correlations between the NEP and DFT are shown in Extended Data Fig. 28. The MD simulations were performed with GPUMD[52]. The initial structure was built by twisting two $\sqrt{626} \times \sqrt{626} \times 5$ STO layers by 4.62°. A 15 Å vacuum slab was applied to the moiré superlattices. The final structure was averaged over the last 25 ps of the total 100 ps trajectory at the temperature of 300 K. The local dipole moment of each unit cell was calculated using the definition[58]:

$$\boldsymbol{P}^i = \frac{1}{\Omega_c}\left(\frac{1}{8}\boldsymbol{Z}_{Sr}^{*}\sum_{k=1}^{8}\boldsymbol{u}_{Sr,k}^{i} + \boldsymbol{Z}_{Ti}^{*}\boldsymbol{u}_{Ti,k}^{i} + \frac{1}{2}\boldsymbol{Z}_{O}^{*}\sum_{k=1}^{6}\boldsymbol{u}_{O}^{i}\right)$$

where $\Omega_c$ is the volume of the unit cell, Z* is the Born effective charge tensor, and u is the atomic displacement from its equilibrium position. The $Sr_2TiO_4$ and $SrTi_2O_5$ structures were used to represent the SrO and $TiO_2$ layers, respectively, in the GSFE calculations.

***Deposition of epitaxial thin films*:** A water-soluble SAO layer was first grown on a $TiO_2$-terminated (001)-oriented STO single-crystal substrate, followed by a thin STO film, both grown by reactive MBE. The thickness of each layer was monitored by RHEED oscillations. The SAO film was grown under an oxygen partial pressure of $1 \times 10^{-6}$ Torr and at a substrate temperature of 850 °C[36]. The STO film was grown under an oxygen partial pressure of $1 \times 10^{-6}$ Torr at a substrate temperature of 700 °C. STO growth was initiated by co-deposition, and for the final several layers, the growth was switched to a shutter-controlled mode. This growth protocol produced high-quality STO films with a well-defined $TiO_2$ surface termination[59].

***Release and transfer of STO films:*** The as-grown 10 × 10 $mm^2$ STO film samples possessed four straight edges, which were used as optical alignment references during the transfer process. The samples were diced into 2.5 × 2.5 $mm^2$ pieces and marked in pairs. During transfer, the reference edges were used to define the initial crystallographic orientation, followed by controlled rotation to achieve the desired relative twist angle between two STO layers.

Prior to transfer, PMMA (A4 950k) was spin-coated onto the STO film surface at 1000 rpm for 45 s. The PMMA-coated side was then attached to a PDMS support layer, and the sample was immersed in deionized water to dissolve the sacrificial layer and release the freestanding STO film. The released freestanding STO film was transferred onto a SiN substrate that had been treated by UV light for 6 min. The film orientation was aligned by matching the reference edge of the STO film with the edge of the substrate. After sufficient conformal contact was established and interfacial stress and trapped bubbles were relaxed, the sample was immersed in 250 mL acetone for 1 h to remove the PMMA support layer.

The second STO layer was transferred using the same procedure. During the second transfer, the reference edge of the upper STO film was aligned with both the reference edge of the first STO layer and the substrate reference edge. The desired twist angle was then introduced using the transfer stage before completing the stacking process. The subsequent PMMA removal procedure was identical to that used for the first layer.

***Annealing process:*** After each acetone cleaning step during the transfer process, the samples were annealed at 200 °C for 2 h under a flowing oxygen atmosphere (oxygen pressure regulated to 0.4 MPa). This treatment removes residual polymer residues and enhances the interfacial interaction, resulting in the rigidly stacked samples shown in Fig. 2b and Fig. 2e. For the formation of the reconstructed interface, the transferred double-layer $SrTiO_3$ samples were heated to 750 °C at a rate of 5 °C $min^{-1}$, maintained at 750 °C for 2 h, and then naturally cooled to room temperature under the same oxygen atmosphere (oxygen pressure regulated to 0.4 MPa). This high-temperature oxygen annealing process induces atomic-scale interfacial reconstruction. A schematic overview of the corresponding procedures is provided in Extended Data Fig. 7.

***Structural characterization:*** The crystal structure was examined by a high-resolution single crystal X-ray diffractometer using a Bruker D8 Discover instrument. Cu $K_{\alpha 1}$ radiation ($\lambda$ = 1.5406 Å) was used.

***Piezoresponse force microscopy measurements:*** The local ferroelectric properties were measured on freestanding films on a conducting Si wafer using an Asylum Research MFP-3D Origin+ SPM scanning probe microscope. Nanoworld EFM-50 Pt/Ir-coated tip (2.8 N $m^{-1}$ in force constant) were used in the PFM measurements. PFM images were taken in the DART mode with driving voltage (2 V a.c.) applied at the tip.

***Scanning transmission electron microscopy (STEM) and electron energy loss spectroscopy (EELS):*** Atomic-resolution STEM imaging was performed on an aberration-corrected FEI Titan Cubed G2 60–300 S/TEM operated at an accelerating voltage of 300 kV, with a probe convergence semi-angle of 22 mrad. HAADF-STEM and ABF-STEM images were acquired at a camera length of 73 mm, corresponding to collection semi-angle ranges of approximately 80–200 mrad and 10–20 mrad, respectively.

STEM–EELS measurements were conducted on a probe-aberration-corrected JEOL JEM-ARM300F microscope (Grand ARM, JEOL Ltd., Tokyo, Japan) operated at 300 kV and equipped with a Gatan K3 direct electron detector. EELS spectrum images were acquired from interfacial regions corresponding to the atomic-resolution HAADF/ABF-STEM images. During acquisition, the beam current, probe convergence semi-angle, energy dispersion and pixel dwell time were set to 50 pA, 22.7 mrad, 0.18 eV per channel, and 10 ms, respectively.

For all EELS datasets, the energy scale was calibrated using the simultaneously acquired zero-loss peak. Prior to quantitative analysis, the pre-edge background was subtracted from the core-loss spectra. The fine structures of the Ti $L_{2,3}$ and O K edges were extracted to evaluate local valence modulation across the interface. Spatial variations in the Ti $L_3$ peak position, the $L_3/L_2$ intensity ratio and the O K-edge spectral features were further correlated with the moiré interfacial structure observed in the HAADF/ABF-STEM images.

***Integrated differential phase-contrast scanning transmission electron microscopy (iDPC-STEM):*** Integrated differential phase contrast (iDPC) STEM was performed on an aberration-corrected microscope (Spectra 300, Thermo Fisher) equipped with the field emission gun and operated at 300 kV. A probe current of 20 pA and a convergence semi-angle of 28.7 mrad were used in HAADF-STEM, whereas the iDPC and DPC images were acquired using a 16-segment detector with a semi-angle range of 7 to 27 mrad.

***Geometric phase analysis (GPA):*** In atomic-resolved HAADF-STEM images, strain analyses were based on Geometric Phase Analysis (GPA) using Gatan DigitalMicrograph software, and the atom positions were accurately determined by using 2D Gaussian peaks fitting in Matlab, which enabled us to acquire the ionic displacement. The lattice rotation is defined as the deviation angle of each lattice unit relative to the ideal crystal axes (horizontal x and vertical y directions), with clockwise and counterclockwise rotations assigned as negative and positive, respectively.

***Scanning microwave impedance microscopy (sMIM):*** Experiments were performed using a Microwave Impedance Microscopy by PrimeNano ScanWave™. The entire system is integrated into an atomic force microscope Attodry 2100. The microwave frequency was approximately 3 GHz. The solid-metal sMIM probes had a tip radius of ~50 nm, enabling simultaneous acquisition of AFM and sMIM signals.
We define $V_{sMIM}$ as the sMIM signal voltage, $V_{in}$ as the input microwave voltage, G as the amplifier gain, and $e^{i\phi}$ as the phase shift of the reference signal. The total admittance of the probe and sample is denoted by $Y_{total}$, which is the sum of the probe admittance $Y_{probe}$ and the complex tip-sample admittance $Y_{ts}$. The relationship is given by:

$$V_{sMIM} = V_{in}Ge^{i\phi}f(Y_{probe}) + V_{in}Ge^{i\phi}f'(Y_{probe})Y_{ts}$$

The existence of the impedance matching network ensures that the probe impedance is substantially matched; consequently, the first term of the aforementioned equation can be approximated as a zero constant. By adjusting the phase shift $\phi$ of the reference signal such that $e^{i\phi}$ becomes a real number, the real and imaginary parts of $V_{sMIM}$ become directly proportional to the real and imaginary parts of $Y_{ts}$, respectively.

***Ethics declaration:*** not applicable.

## Data Availability

The data that support the findings of this study are available from the corresponding author upon reasonable request.

**Acknowledgements**
This work is supported by the National Natural Science Foundation of China (Grant Nos. U24A2011, 52672155, 12434002, 12125404, T2495231, and 12504277), National Basic Research (Key R&D) Program of China (Nos. 2022YFA1402502, 2022YFA1403201), Guangzhou National Laboratory Special Project (GZNL2023A03008), Advanced Materials-National Science and Technology Major Project (Grant 2024ZD0607000), and Basic Research Program of Jiangsu (Grants BK20233001, BK20241253, and BK20253009).

**Author Contributions**
L. Han, S. Cai, J. Sun, D. Wu and Y. Nie conceived the idea and supervised the project. P. Fei led the sample growth, transfer and characterization, performed the PFM measurements, and analysed the key experimental data with assistance from Y. Nie, B. Hao, L. Han and M. Wang. M. Wang supervised the electron microscopy characterization method, sample preparation and data analysis, performed cross-sectional sample preparation and acquired electron microscopy data. S. Liu acquired and analysed additional plan-view electron microscopy data, including GPA and iDPC analyses and *in situ* electron microscopy switching experiments, under the supervision of S. Cai. Z. Zhang carried out the theoretical calculations under the supervision of J. Sun and with the help of J. Wang and C. Ding. H. Zhang and Z. Fei helped with the material transfer methodology. G. Shen, D. Ma, S. Yang, W. Nie, J. Wang, and J. Xu assisted with sample transfer and characterization. H. Chen performed the optical measurements under the supervision of L. Ma. P. Fei, L. Han and Z. Zhang wrote the manuscript. All authors discussed the results and contributed to the manuscript.

**Competing interests**
Authors declare no competing interests.

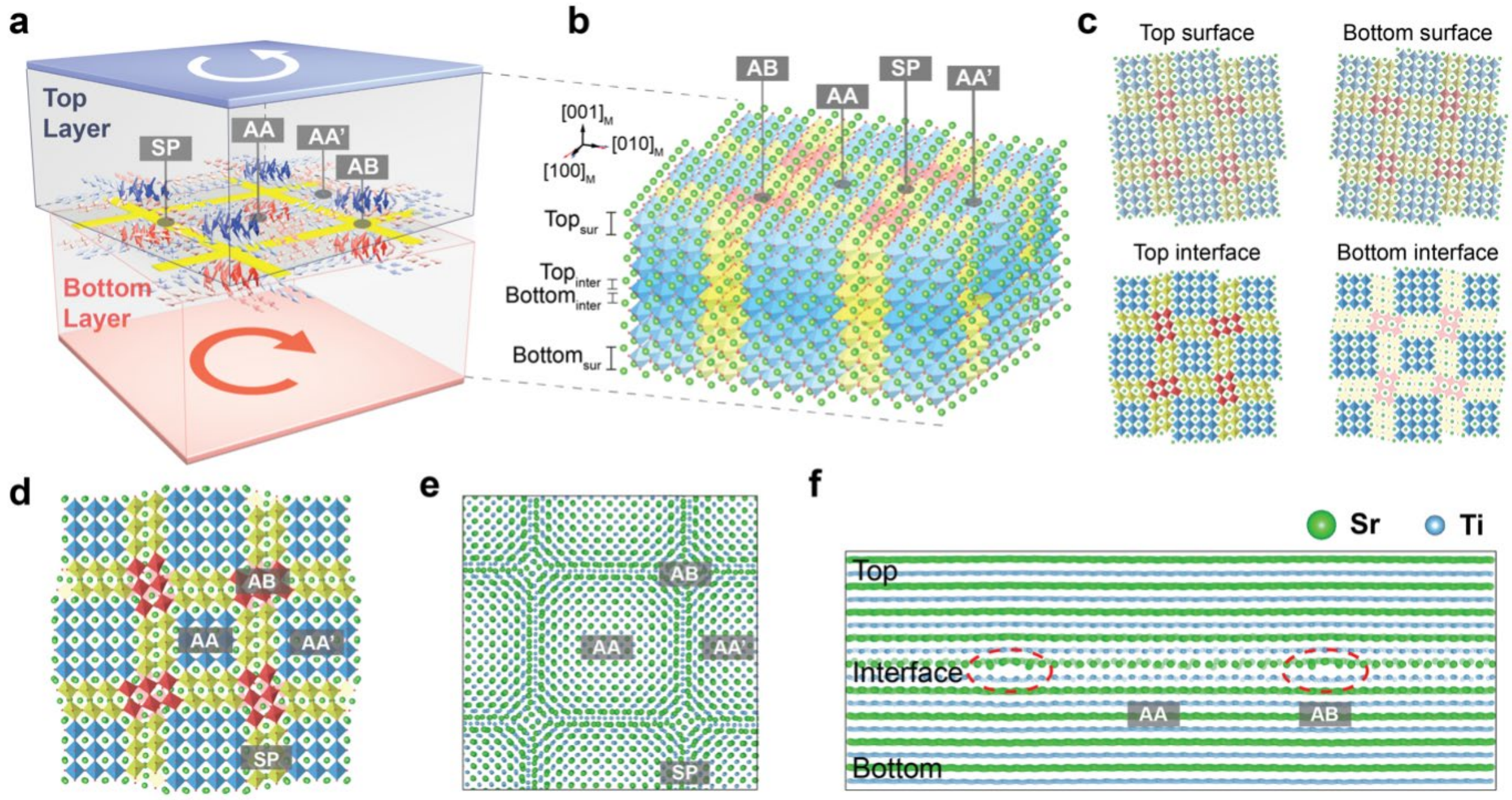

**Fig. 1 | Moiré-modulated structural reconstruction and polar order in twisted $SrTiO_3$ double layers with SrO-$TiO_2$ termination.** **a**, Schematic illustration of the reconstructed structure and polar order at the twisted interface derived from molecular dynamics simulations. Blue and red arrows indicate polar-distortion directions. **b**, **c**, Three-dimensional architecture of the twisted double layer (**b**), comprising four distinct horizontal planes (**c**): top surface ($Top_{sur}$), top-layer interface ($Top_{inter}$), bottom-layer interface ($Bottom_{inter}$), and bottom surface ($Bottom_{sur}$). **d**, Schematic of the relaxed interface structure. **e, f**, Relaxed interface structure of the 4.62° twisted double layer obtained from MLP-based MD simulations: planar view (**e**) and cross-sectional view (**f**).

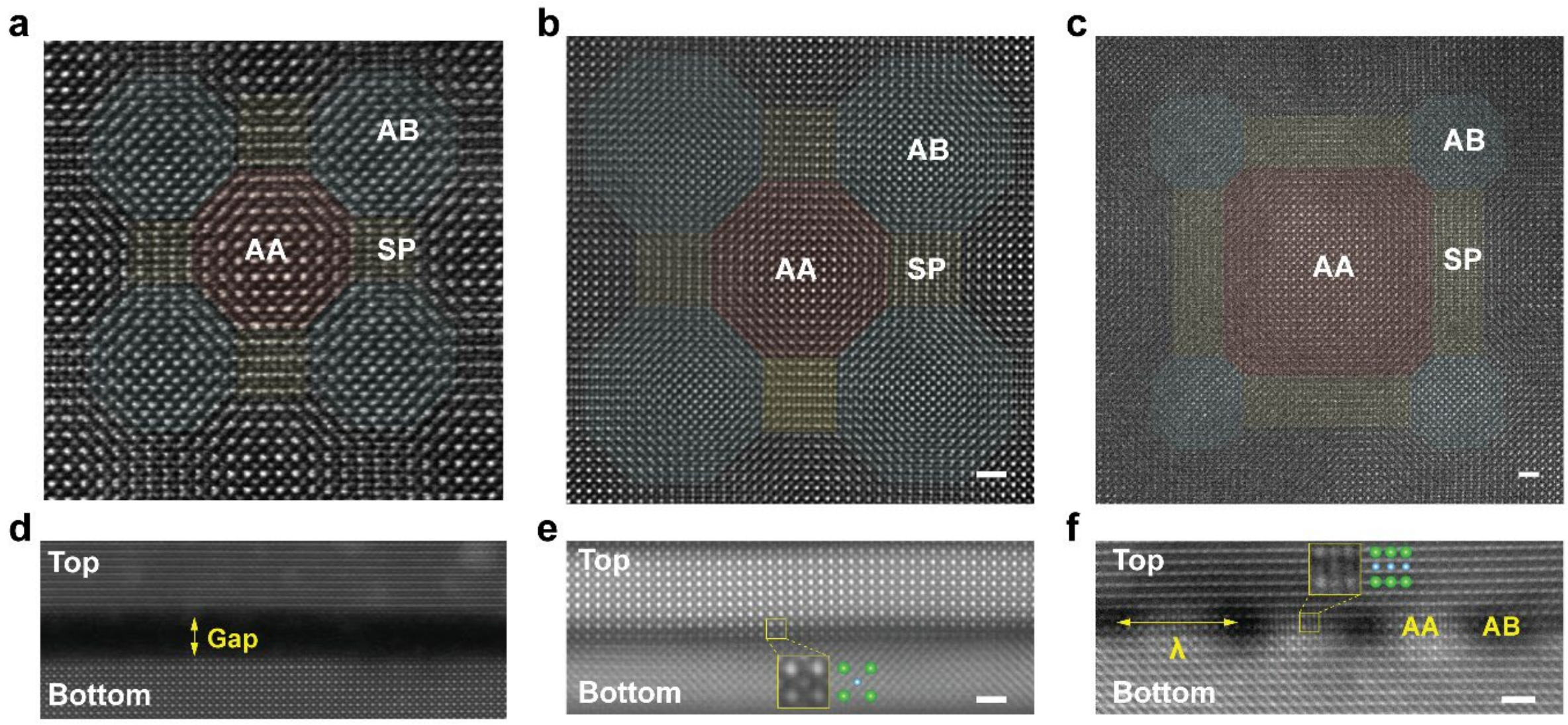

**Fig. 2 | Plan-view and cross-sectional HAADF-STEM images of twisted STO double layers with SrO-TiO$_2$ termination. a**, Plan-view image of a double layer with an amorphous interfacial gap; **d,** the corresponding cross-sectional image. **b**, Plan-view image of a rigidly stacked double layer, showing the atomic structure of an unrelaxed interface; **e**, the corresponding cross-sectional image. **c**, Plan-view image of the reconstructed double layer after annealing, featuring redistributed AA, AB, and SP regions; **f**, the corresponding cross-sectional image highlighting the reconstructed interface. Green and blue circles denote Sr and Ti atomic columns, respectively. Scale bars, 1 nm.

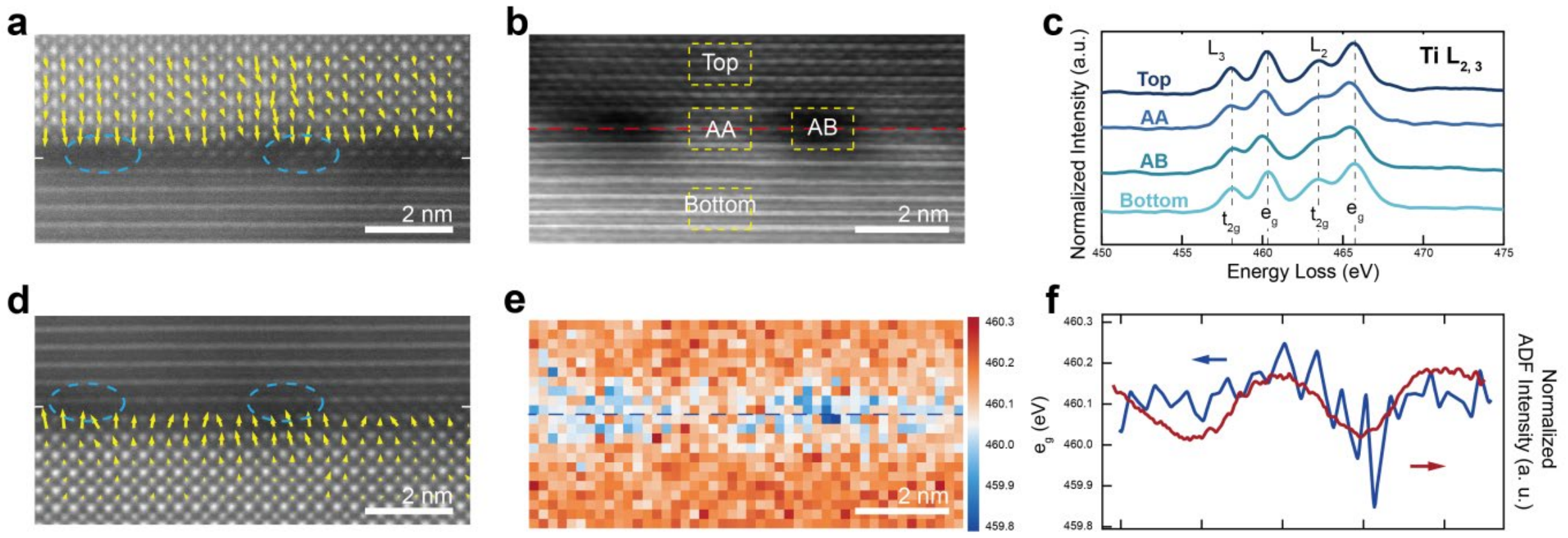

**Fig. 3 | Projected polar-distortion analysis and interfacial EELS mapping. a, d,** Polar-distortion maps at the interface of the top (**a**) and bottom (**d**) layers, respectively. Head-to-head-like polar-distortion is observed near the interface, with a moiré-modulated rotational trend that decays rapidly away from the interface. **b,** Cross-sectional ADF image acquired with the zone axis aligned along the interface. Four equal-area regions, located in the top layer, AA-like region, AB-like region and bottom layer, respectively, were selected for EELS analysis. **c,** The corresponding Ti-$L_{2,3}$ spectra for these four areas. **e,** Two-dimensional mapping of the Ti $L_3$ $e_g$ peak position in the region shown in (**b**), revealing a pronounced red-shift in the moiré AB-like region. **f,** The line profiles extracted from the same position in (**b**) and (**e**), showing a spatial correlation.

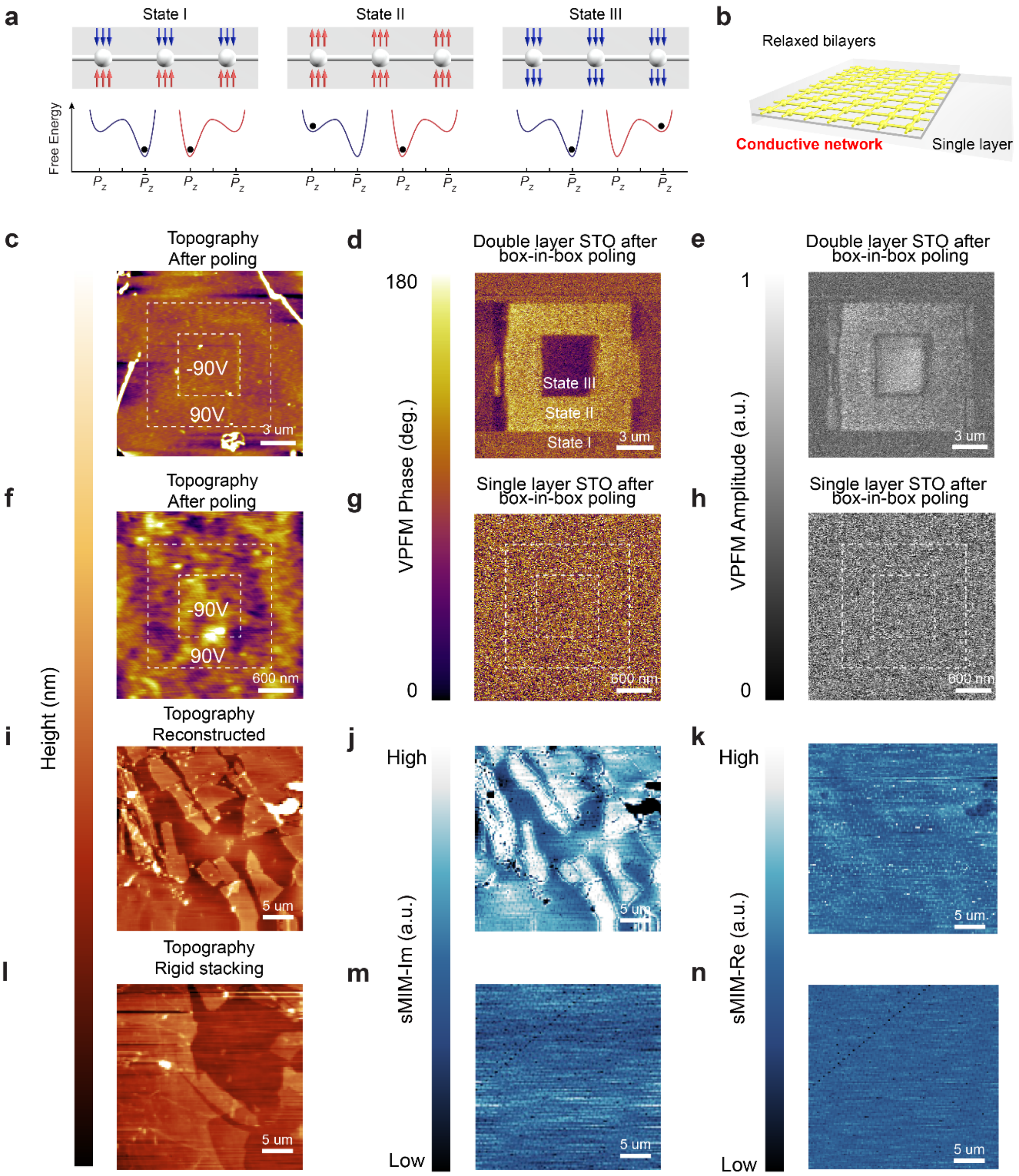

**Fig. 4 | Macroscopic electrical response of twisted STO double layers. a,** Schematic illustration of three-state ferroelectric switching, showing the double-well potential of Landau free energy for AB/SP regions in the top and bottom layer for each state. **b,** Schematic of the buried conductive network at the reconstructed interface in twisted STO double layers. **c, d, e,** Topography(**c**), PFM phase (**d**) and amplitude (**e**) images of the double layer region after box-in-box poling. **f, g, h,** Topography(**f**), PFM phase (**g**) and amplitude (**h**) images of the single layer region on the same sample, it was measured using identical writing and reading protocols and does not show PFM contrast. **i, j, k,** Topography (**i**), imaginary-channel sMIM image (**j**) and real-channel sMIM image (**k**) of the reconstructed double layer sample across the

monolayer/double layer boundary. **l, m, n,** Topography (**l**) imaginary-channel sMIM image (**m**) and real-channel sMIM image (**n**) of an unreconstructed double layer STO control sample with the same thickness.